\documentclass[fleqn,usenatbib,referee]{mnras}

\usepackage{newtxtext,newtxmath}

\usepackage[T1]{fontenc}

\DeclareRobustCommand{\VAN}[3]{#2}
\let\VANthebibliography\thebibliography
\def\thebibliography{\DeclareRobustCommand{\VAN}[3]{##3}\VANthebibliography}

\usepackage{graphicx}	
\usepackage{amsmath}	
\usepackage{supertabular}
\usepackage{epsfig}
\usepackage[utf8x]{inputenc}
\usepackage{supertabular}
\usepackage{rotating}

\title{Pushchino multibeam pulsar search. VII. The results of the timing of 12 slow pulsars} 
\author[S. A. Andrianov et al.]{
	S. A. Andrianov,$^{1}$	
	V. A. Potapov,$^{1}$
	S. A. Tyul'bashev,$^{1}$\thanks{E-mail: serg@prao.ru (SAT)}\\
	$^{1}$ P.N. Lebedev Physical Institute of the Russian Academy of Sciences, Astro Space Center, Pushchino Radio Astronomy Observatory,\\
	Radiotelescopnaya 1a, Moscow reg., Pushchino, 142290, Russia \\
	}

\date{ }

\pubyear{ }

\begin{document}
	\maketitle
	
	\begin{abstract}
We have performed timing of a number of known slow pulsars with poorly known coordinates and parameters of their intrinsic rotation. We used data from the archive of round-the-clock monitoring observations on the third (stationary) beam pattern of the Large Phased Array radio telescope (LPA LPI) at a frequency of 111 MHz, which has an unsatisfactory connection of the local quartz time standards to the reference scale (UTC). To compensate for the resulting errors, we applied an algorithm previously developed by us, which uses Pulsar Timescale as an intermediate reference scale to compute corrections to the pulses Times of Arrival (TOAs) measured by the local clocks and to switch to UTC. Analyzing a ten-year observational data set we substantially refined the rotational and astrometric parameters of 12 pulsars. The spin frequencies $\nu$ and their first derivatives $\dot\nu$ were determined with accuracies of $10^{-10}$ Hz and $10^{-19}$ s$^{-2}$, respectively, which is 5-6 orders of magnitude better than the values quoted in the catalogue. The coordinates are determined with accuracies ranging from units to tens of arcseconds.

\end{abstract}
	
\maketitle 
	
\section {Introduction}

The study of radio pulsars traditionally involves two key tasks: searching for new objects and the subsequent analysis of their observed parameters. When a pulsar candidate is discovered during a search, the next step is its confirmation. This can be achieved either through independent observations on another telescope or by means of timing. Timing not only confirms the presence of the primary characteristic of a pulsar its emission of pulses with a period that remains stable for a long time with high accuracy, but also allows one to determine a number of its physical parameters.

Pulsar timing is the procedure of measuring the Times of Arrival (TOAs) of pulses at an observing point on Earth (Topocentric TOAs) in the local timescale, reducing them to the Barycentre of the Solar System and to Barycentric Time (barycentric TOAs) and creating a model of the pulsar, based on the observed series of TOAs, that is described by a set of parameters with a clear physical meaning. These, first and foremost, include the rotational parameters of the pulsar and its celestial coordinates: the rotation period $P$ and its derivatives $\dot P$, $\ddot P$ (or, alternatively, the intrinsic spin frequency and its derivatives $\nu$, $\dot\nu$, $\ddot\nu$), right ascension and declination and their proper motions $\alpha$, $\delta$, $\mu_\alpha$, $\mu_\delta$, and parallax $\pi$ (\citeauthor{Lorimer2004}, \citeyear{Lorimer2004}; \citeauthor{Doroshenko1990}, \citeyear{Doroshenko1990}).

Depending on the qualitative characteristics of the receiving system (single or multiple observing frequencies, accuracy of TOA determination) and of the observational data set (duration, cadence) or the type of object (isolated or binary pulsar), additional parameters describing the physical properties of the object itself can be included in the model: e.g. Keplerian and relativistic post-Keplerian orbital parameters (\citeauthor{Hulse1975}, \citeyear{Hulse1975}; \citeauthor{Damour1986}, \citeyear{Damour1986}; \citeauthor{Kramer2021}, \citeyear{Kramer2021}) as well as parameters of the interstellar and interplanetary medium so called dispersion measure (DM).

High-precision pulsar timing can also be used to test theories of gravity and to search for gravitational-wave radiation (\citeauthor{Kramer2021}, \citeyear{Kramer2021}; \citeauthor{Taylor1989}, \citeyear{Taylor1989}; \citeauthor{Damour1992}, \citeyear{Damour1992}; \citeauthor{Potapov2003}, \citeyear{Potapov2003}; \citeauthor{Kopeikin2004}, \citeyear{Kopeikin2004}; \citeauthor{Zharov2019}, \citeyear{Zharov2019}; \citeauthor{Arzoumaniam2021}, \citeyear{Arzoumaniam2021}).

Experiments aimed at testing physical theories require extremely high accuracy in the determination of model parameters that cannot be achieved with low-precision TOA measurements. Therefore, such observations must be carried out on radio telescopes that provide high sensitivity, have a time service (atomic frequency standards), and enables observations at high radio frequencies where the delays caused by propagation through the interstellar and interplanetary media are minimal. It is also necessary to have regular observational data sets spanning several years.

For search surveys, on the other hand, the maximum possible sky coverage is important, which can be provided by a multibeam antenna pattern and/or a wide antenna pattern of the radio telescope, together with high instrumental sensitivity, that allows pulsar searches on time intervals of a few minutes. The beamwidth of telescopes used for pulsar searches can range from several arcminutes to a few degrees (see, e.g., \citeauthor{Tyulbashev2016} (\citeyear{Tyulbashev2016}); \citeauthor{Sanidas2019} (\citeyear{Sanidas2019}); \citeauthor{Han2021} (\citeyear{Han2021})). Thus, the requirements for radio telescopes used for precision pulsar timing and for search tasks are poorly compatible.

When a new pulsar or a rotating radio transient (RRAT) is discovered on a telescope with a wide antenna pattern, confirming it on a telescope with a narrow beam becomes problematic. To confirm the discovery and/or to include the object in observing programs, a narrow-beam telescope must scan numerous possible positions of the object within the wide beam of the discovery telescope, which is a labor-intensive and costly procedure. Such systems, for example, are the Large Phased Array (LPA) and the Five-hundred-meter Aperture Spherical Telescope (FAST) (\citeauthor{Tyulbashev2016}, \citeyear{Tyulbashev2016}; \citeauthor{Sanidas2019}, \citeyear{Sanidas2019}; \citeauthor{Han2021}, \citeyear{Han2021}): the LPA beam is $1^\circ \times 0.5^\circ$, whereas that of FAST is $3'$, roughly 200 times smaller in the swept area. In such cases high-precision pulsar timing serves as a natural way to refine the coordinates of the pulsar for its subsequent confirmation.

An additional difficulty in confirming the discovery of a pulsar by other radio telescopes is the low initial accuracy of the determined rotation period, which leads to a loss (smearing) of the pulse during long expositions. In search campaigns this accuracy can be $10^{-3}$--$10^{-5}$ s and corresponds to synchronous folding of the observed pulses within a single observing session (for the LPA this time is typically $\sim 3$--$5$ min). To determine the precise value of the intrinsic rotation period over long time spans the new pulsar must be observed regularly (ideally daily), which is practically impossible in large projects where typical pulsar observations are carried out 1-2 times per month (see, e.g., \citeauthor{Keith2024} (\citeyear{Keith2024})).

At the Pushchino Radio Astronomy Observatory (PRAO ASC LPI) a pulsar and transient search project has been running for more than 10 years (\citeauthor{Tyulbashev2016}, \citeyear{Tyulbashev2016}; \citeauthor{Tyulbashev2018}, \citeyear{Tyulbashev2018}). In blind searches more than 300 pulsars have been discovered through their periodic emission and more than 200 pulsars and RRATs via their single-pulse emission\footnote{\url{https://bsa-analytics.prao.ru/en/project/about/}}. Over 200 of the objects discovered in the searches have low-precision period determinations (3-4 significant digits), and for some RRATs the period has not been determined at all. The new objects also have very low coordinate accuracy due to the wide antenna pattern of the LPA. Thus, confirming the discovery of these pulsars by other radio telescopes is a technically challenging task.

At the same time it is clear that timing allows one to determine pulsar coordinates with an accuracy better than an arcsecond, and $P$ and $\dot P$ with an accuracy certainly better than $10^{-7}$ (certainly sufficient for scheduling timing sessions of slow pulsars). However, the digital recorders on LPA3, where daily recordings are made with 128 beams of the stationary antenna pattern, employ quartz generators that do not provide sufficient accuracy of time reference to the UTC scale, and our early attempts at timing were unsuccessful.

In 2025 successful timing of 24 slow pulsars with well-known parameters using data tied to the local quartz standards was carried out (\citeauthor{Andrianov2025}, \citeyear{Andrianov2025}). The typical accuracy achieved was: rotation period $P$ better than $10^{-8}$–$10^{-9}$ s (sufficient to maintain the pulse phase within $1/2\,P$ over time spans of 3–10 years for a typical pulsar), coordinates $\alpha$, $\delta \sim 10''$. The obtained accuracy is roughly two orders of magnitude worse than the typical accuracy achieved in timing slow pulsars on LPA1 using a specialized pulsar receiver. Nevertheless, this accuracy is sufficient to include the discovered objects in the observing programs of large telescopes without the need to scan all possible combinations of coordinates and periods.

In the present work we present timing results for 12 pulsars from the ATNF catalogue (\citeauthor{Manchester2005}, \citeyear{Manchester2005})\footnote{\url{https://www.atnf.csiro.au/research/pulsar/psrcat/}}, whose periods were previously known with an accuracy no better than the sixth decimal place.

Description of observation and data processing principles are presented in Section 2 of this paper. Results of the paper: new estimations of intrinsic rotation parameters as well as celestial coordinates for 12 pulsars  are presented in Section 3. A short discussion of the results and conclusion follow in Section 4.

\section{Observations and processing} 
\label{Observations}

\subsection{Observations}

We used archival data from the Pushchino Multibeams Pulsar Survey (PUMPS; \citeauthor{Tyulbashev2016} (\citeyear{Tyulbashev2016}); \citeauthor{Tyulbashev2022} (\citeyear{Tyulbashev2022})). This survey was carried out on the LPA3 radio telescope, which has a stationary (immobile) antenna pattern consisting of 128 beams arranged in the plane of the meridian. The duration of a source’s recording is determined by the time it takes to cross the antenna pattern aligned with the celestial meridian due to Earth’s rotation; at the celestial equator this equals approximately 3.5 minutes at the half-power beamwidth. The observational data were recorded by three receivers covering declination ranges from $-9^{\circ}$ to $+21^{\circ}$, from $+21^{\circ}$ to $+42^{\circ}$, and from $+42^{\circ}$ to $+55^{\circ}$, respectively. The central observing frequency was 110.4 MHz, with a bandwidth of 2.5 MHz. The band was split into 32 frequency channels of 78 kHz width; the nominal time sampling interval was 12.4928 ms. The recordings are continuous and daily, except for periods of routine antenna maintenance or force majeure circumstances.

\subsection{Accuracy of the local frequency standard}

Data recording from the antenna is carried out in hourly segments, with the start of each segment synchronized to GPS(TS) time with an accuracy no worse than 2 ms by means of NTP servers. In our observations this is sufficient to realize an external reference UTC scale. The behavior of the local timescale relative to the external reference within a single hour is governed by the accuracy of the quartz oscillator that thus realizing the local timescale. In total, the system contains 16 quartz oscillators -- one oscillator per 8 beams (one module) of the LPA. Paper \citeauthor{Andrianov2025} (\citeyear{Andrianov2025}) shows that the quartz oscillators are unsatisfactory as local frequency standards for timing purposes; nevertheless, the use of an intermediate reference (Pulsar Time) scale enables the transition from the local scale to UTC to be constructed.

Pulsar Time was created by timing of 24 bright pulsars with well-known parameters on LPA3. For each of them, timing residuals (TRs) of TOAs in the local timescale were determined and studied over a long (up to 10 years) interval. Up to seven pulsars were observed within one module, permitting common effects unrelated to the pulsars themselves to be isolated for all pulsars in the module. In the course of studying the TRs series, several peculiarities of the local clock behavior were revealed. These contain both effects that can be successfully compensated (see the detailed discussion in \citeauthor{Andrianov2025} (\citeyear{Andrianov2025}) and fundamental limitations.

Let us briefly list the compensable effects:
\begin{enumerate}
\item \textbf{Periodic quasi-linear drifts in the TRs of pulsar TOAs.} These are associated with an accumulating observation start-time error caused by the deviation of the actual oscillator frequency of a module (the module clock) from its nominal value. Due to hourly synchronization to the external timescale, the TRs caused by this effect is maximal at the end of an hourly recording and minimal at its beginning. Since a pulsar crosses the celestial meridian about four minutes earlier from day to day, the periodicity of this effect is roughly 15 days, and the TR varies from its maximum, when the pulsar lies at the end of the hourly recording, to its minimum. Experimentally it has been shown that between synchronizations the clock rate of the module can be regarded as linear. The effect is accounted for either by introducing a linear correction to the TOAs or by averaging them over 15-day intervals.

\item \textbf{Random variations ("jumps") in TRs caused by a failure of the module clock synchronization procedure with the external scale.} This effect is short-term (a single observation), and as a rule is individual for each pulsar (unless two or more pulsars are observed within the same hourly scan). The effect has the character of a "point disturbance." In most cases it is irremovable and is compensated by simple exclusion of the TOA from the series.

\item \textbf{Polynomial drifts in TRs.} This effect is related to the fact that synchronization with the external time standard was not carried out for a long time (a week, a month), which is typical for early observations. The deviation of the local clock and, consequently, of the TOAs on such data segments can amount to several seconds and exceed the pulsar period. Since the effect is common to all pulsars of the module, it can be removed by subtracting from the TOAs a polynomial function that approximates the TRs drift on the segment.
\end{enumerate}

\begin{figure}
\includegraphics[width=0.9\columnwidth]{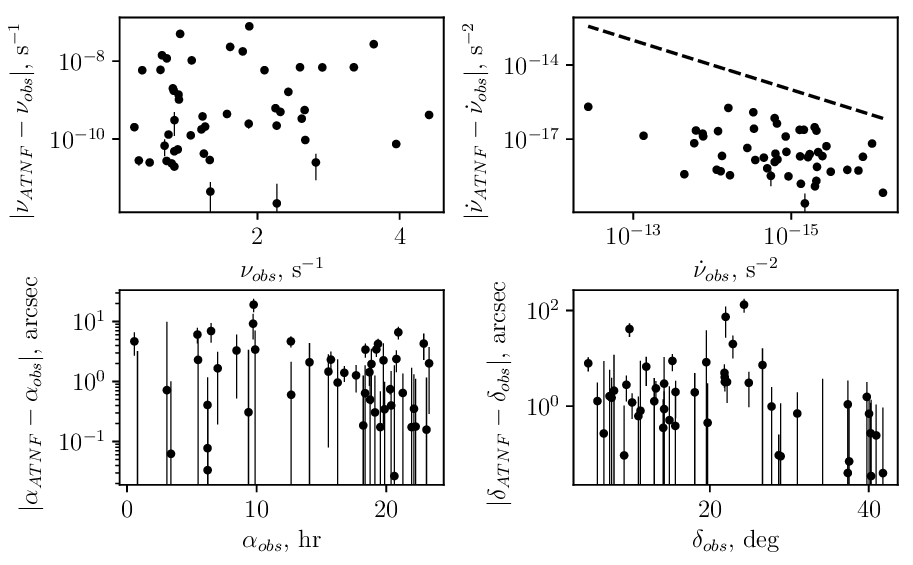}
\caption{Errors in determining the main parameters of the timing model in comparison with the ATNF catalogue. For the plot we postulate that the catalogue parameters are free of errors. The line $x = |y|$ is added to the $\dot{\nu}$ plot: dots, that (potentially) lie above this line, have an estimation of $\dot{\nu}$ that differs from the ATNF's one bigger than their value.}
\label{Fig1}
\end{figure}

Besides these effects, which can be accounted for in timing (\citeauthor{Andrianov2025}, \citeyear{Andrianov2025}), the TRs of pulsars also exhibit a number of effects that cannot yet be compensated:
\begin{enumerate}
\item \textbf{A quasi-sinusoidal component in the TRs of a number of pulsars with a one-year period.} A satisfactory explanation for this effect is still lacking. In the case of observations of pulsars with well-known parameters, this component can be easily removed from the TOAs. When observing pulsars with poorly known coordinates, such a component of TRs can, in some cases, correlate with the orbital R\"{o}mer delay (when its phase is close to the phase of the "orbital sinusoid" in the TOAs) and can lead to a significant distortion of the determined coordinates of the object. The effect is especially strong in the plane of the ecliptic (see Fig.~\ref{Fig1}).

\item \textbf{Pulsar noise.} Model errors of the pulsar rotation are often indistinguishable from clock errors (since both behave as polynomials in time). This becomes critical when fewer than 3–4 pulsars are observed in a module. For reliable compensation of clock drifts it is necessary to observe as many stable reference pulsars as possible.

\item \textbf{Mode switching:} when an observed pulsar exhibits different modes (a jump-like amplification or attenuation of components of the pulse profile, a change in the shape and phase of the pulse). The effect of mode switching leads to an increase in the root-mean-square (RMS) deviations of the TRs and to a decrease in the accuracy of the pulsar parameter determination. However, with a sufficient amount of data the effect can be avoided by using in timing only TOAs, corresponding to a single mode.
\end{enumerate}

In \citeauthor{Andrianov2025} (\citeyear{Andrianov2025}), 24 strong pulsars were selected for the declination range from $+21^{\circ}$ to $+42^{\circ}$, and a technique for extracting the rotational parameters and coordinates of pulsars was developed. Similar work was later carried out for pulsars located between declinations $-9^{\circ}$ and $+21^{\circ}$. On these declinations there are substantially more interferences and the quality of the observations is lower. Nevertheless, we managed to select 25 reference pulsars for this stripe, and for each of them we obtained parameter estimates using our method. Below we list these pulsars: J0034$-$0721, J0304$+$1932, J0525$+$1115, J0659$+$1414, J0922$+$0638, J0946$+$0951, J0953$+$0755, J1404$+$1159, J1543$+$0929, J1614$+$0737, J1645$+$1012, J1740$+$1311, J1823$+$0550, J1841$+$0912, J1844$+$1454, J1851$+$1259, J1913$+$0440, J1932$+$1059, J1946$+$1805, J1952$+$1410, J2037$+$1942, J2046$+$1540, J2116$+$1414, J2215$+$1538, J2253$+$1516.

The rotational and astrometric parameters of these pulsars determined by our method, as well as their TRs before and after clock-drift corrections, are publicly available in the Internet.\footnote{\url{https://bsa-analytics.prao.ru/timing}}

The differences between the parameters derived by us from the LPA3 survey and the values given in the ATNF catalogue for 49 pulsars are shown in Fig.~\ref{Fig1}. We note that the largest discrepancy in the estimates of the spin frequency are obtained when comparing with the results presented in \citeauthor{Hobbs2004} (\citeyear{Hobbs2004}), where observations span more than 30 years. The largest difference in the estimates of the spin-frequency derivative occurs when comparing with \citeauthor{Keith2024} (\citeyear{Keith2024}), where the observation duration does not exceed 3 years. Here both the effect of pulsar noise and the incomplete compensation of our reference clock drifts become evident. We emphasize that the figures show the absolute values of the parameter differences and that no statistically significant systematic shifts in our estimates have been detected.

Over the ten-year data span, the values of $\nu$ and $\dot\nu$ obtained for known slow pulsars from our analysis may differ from the estimates listed in the ATNF catalogue by amounts $\sigma_{\nu\ ATNF}\approx|\nu_{ATNF}-\nu_{obs}|<10^{5}\sigma_{\nu\ ATNF}$ (for the multi-year, multi-frequency observations given in the ATNF catalogue) and $\sigma_{\nu\ 111}\approx|\nu_{111}-\nu_{obs}|<10^{3}\sigma_{\nu\ 111}$ for the single-frequency ($\sim$111 MHz) observations obtained with reliable local time realization, where $\sigma_{\nu\ \dots}$ denotes the least-squares timing parameter error estimation.

The accuracy of the coordinate determinations in right ascension and declination has the largest errors in the ecliptic plane, the reasons for which were described above (the "outlier" in $\delta$ in Fig.~\ref{Fig1} at declinations near $23.5^{\circ}$). The difference between the coordinates derived from the LPA3 data and those obtained in earlier works ranges from $0.1''$ up to $\sim100''$ (for example, J0528$+$2200, which lies almost in the ecliptic plane).

Thus, at present the clock-drift corrections are known for pulsars with declinations from $-9^{\circ}$ to $+42^{\circ}$. They can be employed to perform timing of pulsars whose periods and coordinates are known with low precision.

We selected 12 pulsars from the ATNF catalogue whose periods were known with an accuracy no better than the sixth decimal place (the fifth for one pulsar) to refine their timing solutions. The period derivative was known for 11 of the 12 pulsars. For these pulsars we obtained timing solutions and examined possible problems that may arise when timing relatively faint sources that are only sporadically detected in the LPA3 data.

\subsection{Pulsar timing algorithm}

The algorithm for timing with poor local time standard data is common and was described in \citeauthor{Andrianov2025} (\citeyear{Andrianov2025}). It consists of the following stages:
\begin{enumerate}
\item \textbf{Determination of local TOAs of the pulsar in the local timescale} as the sum of the session start time and the fiducial point within the pulsar period. The integrated pulse profile is the summed (average) profile of the pulsar over one observing session. In this work the fiducial point is taken to be the position of the maximum of the cross-correlation function between the integrated and the template profile. The template profile is the integrated profile, averaged over many observing sessions. The extraction of the integrated and template profiles and the TOAs was carried out with the AnTi-pipeline program.\footnote{\url{https://github.com/StepanRepo/AnTi-pipeline.git}}

\item \textbf{Conversion of the TOAs from the local timescale to UTC.} Such a transition is usually performed by comparing the observatory’s atomic clock with the international standard. In our case, when the local timescale is represented by 16 quartz oscillators that are irregularly compared with GPS(TS), the comparison with UTC is done through an intermediate Pulsar-Timescale (see (\citeauthor{Andrianov2025}, \citeyear{Andrianov2025}) for details).

\item \textbf{Transformation of TOAs to the Barycentric Reference Frame and to Barycentric Time.} This procedure is described in \citeauthor{Doroshenko1990} (\citeyear{Doroshenko1990}); \citeauthor{Hobbs2006} (\citeyear{Hobbs2006}) and is implemented in the programs TEMPO2 and TIMAPR, respectively.

\item \textbf{Computation of modeled TOAs in the Barycentric Reference Frame} using the given pulsar ephemeris, and the determination of timing residuals (TRs) the differences between the modeled and observed TOAs.

\item \textbf{Refinement of the pulsar's model parameters} by minimizing the RMS of the TRs with an iterative least-squares procedure.
\end{enumerate}

Among the pulsars, selected for the timing, there may be both sufficiently bright and weak objects whose emission is only occasionally registered. For low signal-to-noise ratios (S/N), besides the real average profiles of the pulsar, spurious (false) profiles unrelated to the pulsar emission can be accumulated by chance.

The initial template profile is accumulated by summing the integrated profiles in which the observed burst is assumed to belong to the pulsar (is not noise). The profiles are selected manually. As a rule, if the number of false profiles is small, they are easily identified on the plot of TRs versus time and can be removed from further consideration. When necessary (a vast of false profiles appear in the raw data) the template can be re-accumulated.

\begin{figure}
	\includegraphics[width=0.9\columnwidth]{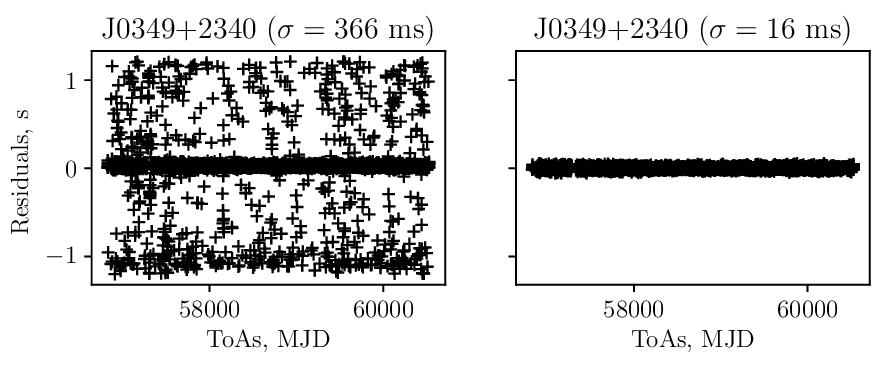}
	\caption{The left panel shows all the TR points obtained for pulsar J0349+2340. The right panel shows at the same scale the TRs after cleaning from spurious pulses and correcting for the quartz clock behavior. The vertical axis represents TRs in seconds and the horizontal axes represent time in Modified Julian Date (MJD).}
	\label{Fig2}
\end{figure}

Fig.~\ref{Fig2} shows the TRs for the relatively bright pulsar J0349$+$2340, discovered by LOFAR in 2019 (\citeauthor{Sanidas2019}, \citeyear{Sanidas2019}) and independently with LPA in 2020 (\citeauthor{Tyulbashev2020}, \citeyear{Tyulbashev2020}). The left-hand panel clearly shows a concentration of TRs within $\pm 20$ ms. There are 3449 points in total on the left panel; after iterative cleaning, 2757 points remain on the right panel. The remaining points belong to spurious peaks. A convenient rejection algorithm is sigma clipping, consisting of iterative determination of the row's standard deviation and exclusion of the points, lying beyond a threshold deviation (e.g., $3\sigma$). The dimmer the pulsar, the less distinct the cluster of points associated with the pulsar emission becomes. In the extreme case the pulsar signal is lost among noise peaks. Timing in this case becomes impossible.

Note that obtaining the integrated profile of the pulsar is a process that is not demanding on the parameters' accuracy; it only requires the clock rate to be accurate enough to avoid ''smearing'' the profile during synchronous summation. The rule is that the phase difference of the profile at the pulse over the length of a synchronous integration session must not exceed $1/2\,W_{0.5}$, i.e., half the width of the integrated profile at 50\% intensity during the session. For a typical slow-pulsar profile width this corresponds to a profile shift of one time sample in the recording over the session ($\approx$12.5 ms), which in turn corresponds to an accuracy in the assumed period $P\sim10^{-5}$ s.

Because the computation of TRs assumes phase coherence over long time intervals within half of a period, refining the period from preliminary estimates, known with low accuracy (3-5 decimal places), can be complicated by the so-called \emph{resonance error}. Such an error leads to TRs separated by an integer number of pulse phase cycles (e.g., an error of one or more pulsar rotations per day). The probability of obtaining such an error increases for weak objects that appear in the observations sporadically. We check for this error by searching for a characteristic TRs' pattern in the data of each observing module that is associated with the behavior of the local standard.

Nevertheless, the resonance error can become significant, especially when obtaining timing solutions for RRATs. For some of them, despite dozens or even hundreds of detected pulses, no reliable period estimate exists. A resonance error can yield a set of periods differing in the fourth or fifth decimal place for RRATs with periods of order of 1 s.

As an example of the resonance error, we consider the timing of the pulsar J1951$+$28 using LPA3 data (Smirnova et al., RAA submitted). This object, with period $P=7.3342$ s ($\nu=0.13635$ s$^{-1}$), was discovered in a search for pulsars with $P>2$ s (\citeauthor{Tyulbashev2024}, \citeyear{Tyulbashev2024}). It has isolated strong pulses and is similar to RRAT. According to the original paper, the period determination error was $0.001$ s. With such an error, between daily observing sessions the number of elapsed periods can lie between 11\,746.7 ($P=7.3352$ s) and 11\,749.9 ($P=7.3332$ s). We list in Tab.~\ref{ResonantError} several parameter sets that arise during timing. The first and second columns give the rotational parameters; the third and fourth columns give the pulsar coordinates. 
\begin{table}
\caption{Rotating and astrometric parameters of pulsar J1951+28 at resonance error}
\begin{tabular}{cccc}
\hline
$\nu$, s$^{-1}$ & $\dot \nu$, $\times 10^{-16}$s$^{-2}$ & $\alpha$ & $\delta$\\
\hline
0.13635308717(2) & -5.4506(12) & 19:51:38(2) & +28:37:42(27)\\
0.13636469293(2) & -5.4507(12) & 19:51:37(2) & +28:37:57(27)\\
0.13638790454(2) & -5.4508(12) & 19:51:37(2) & +28:38:27(27)\\
\end{tabular}
\label{ResonantError}
\end{table}    

The obtained values show that, despite the poor period estimates, the period derivative and coordinates differ only slightly. This allows the parameter estimates to be used in $P/\dot P$ diagrams and to include the new objects in research programs on other telescopes.

In the paper (Smirnova et al., RAA submited) only one set of $P$, $\dot P$ and coordinates is given. This is because, in addition to the LPA3 observations, observations of J1951$+$28 with FAST were also available. In two FAST sessions the pulsar coordinates were determined with an error of less than $1.5'$, and accurate (atomic-clock based) time tags for 5 pulses were obtained. Including the FAST pulse TOAs allowed an unambiguous choice among the sets of $P$, $\dot P$ and coordinates.

Another parameter, used in obtaining the average profile is the dispersion measure (DM). LPA has a narrow receiving frequency band, so the typical accuracy of DM determination is $\sim$1 pc cm$^{-3}$. An incorrect DM value leads to a smearing average pulse profile and hence to degraded accuracy of its TOA determination. In this paper, when refining the pulsar parameters, the DM values, obtained from multi-frequency observations, were taken from the ATNF catalogue and assumed constant over the entire observation interval.

\section{Refinement of timing parameters for 12 pulsars}

\begin{figure}
	\includegraphics[width=0.7\columnwidth]{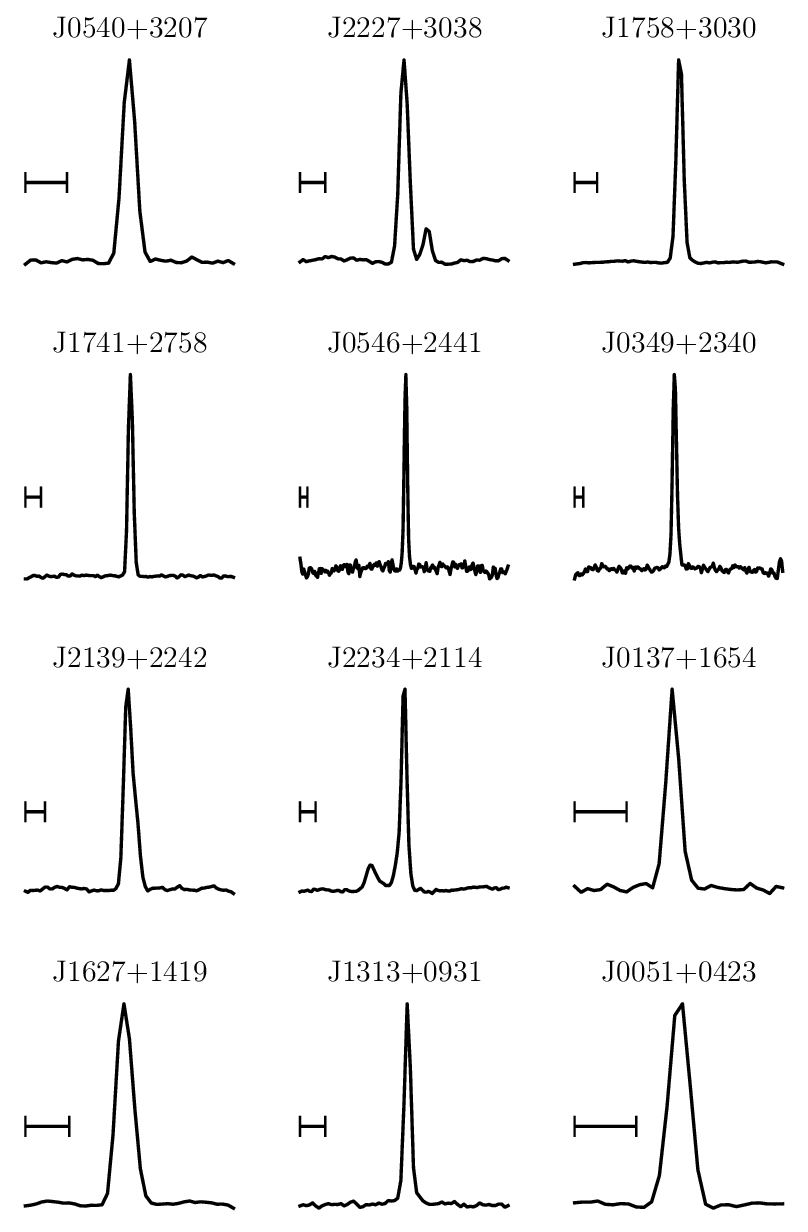}
	\caption{Mean profile of pulsars, obtained by averaging a series of integral profiles. Each plot is labeled with J2000 pulsar name. Fiducial segments on the plots have lengths 100 ms.}
	\label{Fig3}
\end{figure}

Using the algorithm described above and in \citeauthor{Andrianov2025} (\citeyear{Andrianov2025}), we obtained refined parameter values for 12 known pulsars whose periods, according to the ATNF catalogue, had previously been determined with an accuracy no better than $10^{-6}$ s. The results for all pulsars, referred to the epoch MJD 51544, are collected in Tab.~\ref{Parameters}. Column~1 gives the pulsar name in the J2000 convention, columns 2-5 present the estimations of rotational parameters: $\nu$, $\dot\nu$, $P$, $\dot P$. Columns 6-7 give the pulsar coordinates in right ascension and declination at epoch J2000. Column~8 lists the half-power width ($W_{0.5}$) of the mean profile. According to \citeauthor{Andrianov2025} (\citeyear{Andrianov2025}), the true half-width may be smaller by the sampling interval (12.5 ms).

The uncertainties are given in parentheses next to the last significant digit. For columns 2-5 the quoted error is the 1$\sigma$ least-squares estimate; for the coordinates the error is 10$\sigma$ (to account for the potential systematic error arising from a quasi-sinusoidal drift in the TOAs of some pulsars on LPA3, which can affect the coordinate estimate; see Sec.~\ref{Observations} for details). Parameter estimations and their uncertainties were obtained with the program TEMPO2. Resonant errors were excluded because the achieved period accuracy allows the pulse phase to be tracked between successive LPA3 observing sessions.

\begin{table}
		\caption{Refined estimates of rotational parameters and coordinates of 12 pulsars}
		\centering
\begin{tabular}{cccccccc}
	\hline
Name & $\nu$, s$^{-1}$ & $\dot \nu$, $10^{-16}$s$^{-2}$ & $P$, s & $\dot P$, $10^{-16}$ s/s & $\alpha$, hh:mm:ss & $\delta$, y:m:d  & $W_{0.5}$, ms \\
\hline
	J0540+3207 & $1.9074114344 (6)$ & $-16.312 (8)$ & $0.5242707378 (2)$ & $59.347 (2)$ & $05:40:37.1 (1)$ & $+32:07:48 (8)$ & $35$ \\
	J2227+3038 & $1.18707335619 (4)$ & $-16.5227 (6)$ & $0.84240792263 (3)$ & $23.2829 (4)$ & $22:27:41.7 (1)$ & $+30:38:22 (2)$ & $40$ \\
	J1758+3030 &$1.05568092131 (2)$ &$-7.8821 (3)$ &$0.94725591778 (2)$ &$8.7843 (3)$ &$17:58:25.86 (5)$ &$+30:30:23 (1)$ &$34$ \\
	J1741+2758 &$0.73489551535 (6)$ &$-9.8223 (9)$ &$1.3607376547 (1)$ &$5.305 (2)$ &$17:41:53.5 (2)$ &$+27:58:09 (9)$ &$37$ \\
	J0546+2441 &$0.35163605083 (2)$ &$-9.4749 (3)$ &$2.84384947916 (14)$ &$1.1716 (23)$ &$05:46:28.5 (2)$ &$+24:42:37 (80)$ &$47$ \\
	J0349+2340 &$0.41309164292 (2)$ &$-1.8670 (3)$ &$2.4207703475 (1)$ &$0.319 (2)$ &$03:49:56.9 (5)$ &$+23:40:50 (30)$ &$54$ \\
	J2139+2242 &$0.92292324025 (2)$ &$-12.1140 (3)$ &$1.08351372724 (2)$ &$10.3186 (3)$ &$21:39:20.0 (1)$ &$+22:42:37 (2)$ &$53$ \\
	J2234+2114 &$0.73597307464 (2)$ &$-1.1905 (3)$ &$1.35874535966 (4)$ &$0.6448 (6)$ &$22:34:56.74 (9)$ &$+21:14:13 (2)$ &$40$ \\
	J0137+1654 &$2.4110152961 (1)$ &$-0.692 (2)$ &$0.41476302602 (2)$ &$4.0212 (3)$ &$01:37:22.9 (4)$ &$+16:55:20 (10)$ &$27$ \\
	J1627+1419 &$2.03725396679 (3)$ &$-16.2941 (5)$ &$0.49085681820 (1)$ &$67.6271 (1)$ &$16:27:18.66 (4)$ &$+14:19:17 (1)$ &$41$ \\
	J1313+0931 &$1.17794955426 (3)$ &$-10.5291 (5)$ &$0.84893278866 (2)$ &$14.6099 (4)$ &$13:13:23.1 (1)$ &$+09:31:53 (3)$ &$29$ \\
	J0051+0423 &$2.81903116420 (7)$ &$-0.662 (1)$ &$0.35473180031 (1)$ &$5.2572 (1)$ &$00:51:30 (1)$ &$+04:22:60 (50)$ &$36$ \\
\end{tabular}
\end{table}    
\label{Parameters}

Fig.~\ref{Fig3} shows the mean pulse profiles of the pulsars, obtained using the derived timing solutions. For J2227$+$3038 and J2234$+$2114 the post-cursor and precursor components are clearly visible in the profiles. We verified their existence against earlier publications. The observed profile details are confirmed at much lower S/N in \citeauthor{Sobey2022} (\citeyear{Sobey2022}) (LOFAR; 144 MHz) and \citeauthor{Deneva2024} (\citeyear{Deneva2024}) (Arecibo; 327 MHz).

\section{Discussion and conclusion}

Of the 12 slow pulsars, presented in Tab.~\ref{Parameters}, 11 had a period estimate in the catalogue known to the 6th decimal place, and one to the 5th decimal place. Likewise, 11 pulsars had a period-derivative estimate quoted to 3 significant digits, while one object had no $\dot P$ estimate at all. As a result of analyzing the LPA3 data we obtained spin-frequency and spin-frequency-derivative estimates with accuracies of about $10^{-10}$~s$^{-1}$ and $10^{-19}$~s$^{-2}$, respectively (corresponding to $10^{-10}$~s and $10^{-19}$~s${}/{}$s for the period and its derivative). For 7 objects the derived coordinate estimate agrees within 1$\sigma$ with the ATNF values. The coordinate estimates for 10 objects agree within the stated 10$\sigma$ uncertainty, which in absolute terms ranges from 1 to 90 arcseconds.

According to \citeauthor{Andrianov2025} (\citeyear{Andrianov2025}), a "soft criterion" was adopted in the timing process: a timing solution was deemed successful if the integrated pulse did not smear by more than half a period over the entire interval of observations. In the same study, using the example of pulsar J0528$+$2200, it was shown that by employing a smaller number of observing sessions one can  achieve only a small broadening of the average profile, of the order of one sample.

Consequently, one might obtain a slightly smeared average profile by coherently summing a number of sessions. The quality of the resulting profile will be lower than in standard pulsar observations due to the broadening: the narrower the profile or its components, the less detail will be visible. Nevertheless, even such a profile can provide additional information for poorly studied pulsars for which only discovery papers were published. In particular, interpulses, precursors, postcursors and other features may be detected in these profiles (\citeauthor{Smirnova2024}, \citeyear{Smirnova2024}; \citeauthor{Toropov2024}, \citeyear{Toropov2024}). Two such pulsars are shown in Fig.~\ref{Fig3} (J2227$+$3038 and J2234$+$2114).

The accuracy achieved for the pulsar coordinates and rotational parameters is typically two orders of magnitude worse than that of standard timing, in which atomic frequency standards and/or reliable GPS time transfer are used to control the clock. However, the accuracy we obtain is more than sufficient for confirming newly discovered pulsars and for scheduling follow-up observations on other radio telescopes.

In conclusion, we summarize the main results of the work:
\begin{itemize}
\item Timing of a sample of 25 known bright pulsars with well-determined parameters has been carried out to obtain clock-drift corrections for the declination stripe $-9^{\circ}<\delta<+21^{\circ}$.

\item Limitations of pulsar timing with a local frequency standard of inadequate accuracy have been examined.

\item Using the timing method, estimates of the rotational and astrometric parameters of 12 known pulsars have been obtained; these pulsars previously had unsatisfactory estimates of their spin frequencies and their derivatives.
\end{itemize}

\section*{Acknowledgements}

The authors are grateful to the antenna group, maintaining LPA, for continuos monitoring of its performance and repair. We also thank L.B.~Potapova for her help in preparing this manuscript.

\end{document}